\documentclass{scrartcl}

\usepackage[utf8]{inputenc}
\usepackage{lmodern}
\usepackage[english]{babel}
\usepackage{amsmath,amssymb,amsthm}

\usepackage{algorithm}

\title{Notes on socio-economic transparency mechanisms}
\author{Philipp Geiger\\mail@philippgeiger.org}
\date{}

\begin{document}

\maketitle

\begin{abstract}
\noindent Clearly, socio-economic freedom requires some extent of transparency regarding the implications of choices.
In this paper, we review some established mechanisms for achieving such transparency, without any claim to completeness, and briefly discuss potential future directions.
Our investigation is structured by four ``challenges'' under which we subsume the various requirements on, and approaches to, socio-economic transparency mechanisms.
One main focus is on the inference, i.e., statistical, aspect of such mechanisms.
\end{abstract}

\section{Introduction}

Is it freedom if one can chose between options, but does not know their implications?
Obviously not.
For actual freedom it is necessary that implications are at least somewhat transparent \cite{wiki:Information_economics}.

Many ``mechanisms'' have been established to make socio-economic choices more transparent:
Classical approaches are  publicly available product tests by consumer safety groups, nutrition information printed on food products (sometimes enforced by official regulations), or simple labels, established by governmental or non-governmental organizations, which indicate certain ``organic'' production methods or ``fair'' salary conditions for producers.
Modern, digital forms of such mechanisms comprise consumer ratings and reviews on web shops like amazon.com, which proved very successful and show the potential of modern information technology to address the problem of transparency of choices.
Another example are salary comparison web sites like glassdoor.com.
In the field of politics, another type of digital mechanism has been established in recent years: platforms, such as wikileaks.org, that take reports or documents, e.g., on military incidents, as input and make some of them available to the public, in particular to allow more informed election decisions. 
Here, anonymity is a major challenge.

Regarding other socio-economic choices, questions about their implications remain unanswered or are only insufficiently answered, for instance:
``What was the average/ median/ minimum salary of the workers that produced this blue jeans I am thinking about buying?'',
``Which chemicals were used in its production, and how do they affect the environment?'',
``How long will this washing machine I am thinking about buying work until it breaks, under normal usage?'', or
``Does it make sense to invest time in applying for this job vacancy, or are there too many other competitors?''

The goal of this paper is to lay some ground for potential future research in the direction of socio-economic transparency instruments, in particular from the perspective of \emph{inference}.
To this end, we first present a preliminary definition in Section \ref{sec:pre}.
Then, in Section \ref{sec:cha}, we try to elucidate the various challenges w.r.t.\ socio-economic transparency mechanisms by subsuming them under four preliminary categories.
For each category, we discuss some established approaches to these challenges (mainly from practice), and mention some ideas for future directions.
We conclude with Section \ref{sec:conclusion}.

It needs to be emphasized that we do not claim any significant original contribution in this paper and, due to the breadth of the topic, postpone a detailed review of related \emph{scientific} work (from economics, sociology, political science, computer science, statistics, machine learning, or philosophy) to future work. 
%

\section{Approximate definition of socio-economic transparency mechanism}
\label{sec:pre}

What we refer to as ``(socio-economic) transparency mechanism'', and what may have different names elsewhere, can be characterized by the following necessary (certainly not sufficient) conditions:
\begin{itemize}
\item It provides an instrument that either allows to search the information base for pieces that are relevant for a given, specific question related to a socio-economical choice, or, ideally, directly answers them.
We sometimes refer to (the usage of) this instrument as ``inquiry'' and the person who asks the question the ``inquirer''.
\item It provides an instrument to enter socio-economic information.
We sometimes refer to (the usage of) this instrument as ``reporting'', the input as ``report'', and the person as ``reporter''.
\end{itemize}

Further necessary characteristics will be established in Section \ref{sec:cha}.

\section{Challenges for the design of transparency mechanisms}
\label{sec:cha}

Here, we identify four key challenges, or requirements, for the design of socio-economic transparency mechanisms:
\begin{itemize}
\item language challenge: discussed in Section \ref{sec:cha_language},
\item inference challenge: discussed in Section \ref{sec:cha_inference},
\item truth challenge: discussed in Section \ref{sec:cha_truth},
\item privacy challenge: discussed in Section \ref{sec:cha_privacy}.
\end{itemize}
Clearly, all these challenges are intertwined.

\subsection{Notes on the language challenge}
\label{sec:cha_language}

For reports to be connectible to potential questions, it is necessary that both, reports and questions, are formulated in (not necessarily the same) standardized language:
the ``form'' of a report alone is not worth anything.
At least there has to be consensus on which entities names in reports (e.g., company names) refer to in reality.

This language needs to be established, and a major challenge is to neither make it too narrow, because then information cannot be transported well enough by it, nor too expressive, because then inference gets difficult.
The most extreme case of a narrow language would be to have just a single dimension (say a real number), think of a salary, which would make inference using statistic, machine learning or related methods rather easy.

\subsection{Notes on the inference challenge}
\label{sec:cha_inference}

The inference challenge can be described as follows:
in principle, a transparency mechanism could simply make all reports entered so far available in an unstructured data base - which certainly is a ``sufficient statistic'' of all available information.
However, besides privacy problems, depending on the form of the input, it can be tedious to practically impossible to filter out the information relevant to answer a specific question.

In some sense, the internet\footnote{In the future, an even more general version could be the internet of things in the future, in the sense that more \emph{sensory} data would be made available.}
can be seen as the most general form of such a ``non-inferring'' mechanism.
(Note that, being not explicitly handled as a ``transparency mechanism'', e.g., in terms of privacy, the internet also may discourage important sources to input their information.)

On the other end of the spectrum, there are consumer rating mechanisms (e.g., amazon.com) or salary databases (e.g., glassdoor.com).
While these mechanisms are quite limited in their scope and input, this limitations also brings about the great advantage of allowing automated \emph{inference} of ``summary'' answers from the available information:
e.g. on amazon.com, one can query the intuitively accessible average number of ``stars'' that are used to rate products, as well as the ``sample size'' of ratings;
on glassdoor.com, besides averages, other statistics such as minimum or maximum are (apparently) provided as well.

A powerful summary of \emph{reviews} can be observed on amazon.com again: one can compare the best against the worst review, a ``statistic'' which is also very relevant for the truth challenge discussed below.

Besides the problem of summarizing the available data, we also consider the classical problem of induction \cite{hume1955inquiry} (extrapolation) as part of the inference challenge: e.g., how relevant is a salary entered three years ago for judging how a company currently treats its employees?
Or: what does it say about the lifetime of this washing machine when the distribution of lifetimes of other machines of the same type was like this and this.
All in all, it seems promising to try more sophisticated methods from machine learning and natural language processing to address the automatable aspects of this challenge.

\subsection{Notes on the truth/gaming challenge}
\label{sec:cha_truth}

Clearly there is a spectrum of types of information that are relevant for transparency, reaching from objective, falsifiable, reproducible facts (say the salary after taxes a person received in a specific month from a specific employer) to more subjective judgments (say the emotional satisfaction with a certain product).
Here we rather focus on the first kind of information, the kind which can be ``correct'' (not to say ``true'') versus ``incorrect''.

For whatever question an inquirer inputs into the transparency mechanism, he/she obviously desires a \emph{correct} answer, or at least correct hints - or an output indicating that no relevant information could be retrieved.
Even in case the inference component of a mechanism is sound - consider the simple case of the mean salary on glassdoor.com actually calculating the empirical average, a consistent estimate - the correctness of any answer it gives obviously depends on whether the entered information is correct.
Obviously, such correctness cannot be assumed in general. People may undeliberately enter wrong information, sensors may break and deliver wrong data.
But in particular, various people or organizations have various interests w.r.t.\ the output behavior of the mechanism and will thus enter all kinds of wrong or misleading information that leads to the desired output, thus ``gaming'' the mechanism.
This is a serious problem for transparency mechanisms and no general solution, i.e., one which does not make significant assumptions on the population of reporters, seems possible.

However, simple ideas have been developed that are surprisingly powerful tools against some extent of gaming.
For instance, in case of one-dimensional ratings of products, such a tool consists of the statistic (a pair) comprising the minimum and the maximum rating.
Assume that a product has many overly positive fake ratings and only one honest negative rating: then this statistic will still output the one honest rating (together with one fake rating).
Note that a slightly more sophisticated version of this idea is implemented on amazon.com.

To paraphrase the basic idea: \emph{if there is a discrepancy in information, this discrepancy needs to be displayed}.
This makes sure that, whenever there are at least some truthful players, some correct information will reach the inquirer.
A version of this idea is also very successfully implemented in the form of wikipedia.org discussions and version histories.

More generally, for the case of online reviews, first steps have been taken for applying machine learning to reduce ``review spam'' \cite{crawford2015survey}, mainly based on supervised learning techniques.

Other approaches may be based on the definition of ``knowledge'' by Plato \cite{campbell1883theaetetus}: it is not correct knowledge if you do not have an argument or some form of evidence.
To test evidence provided by some reporter, each element of the provided evidence can be compared to known facts, or to information provided by other reporters.

Yet another direction may be based on ``trustworthiness graphs'': 
one could imagine that every person writes down a trustworthiness value for a collection of other persons, which forms a weighted graph.
This could serve as one additional input to trustworthiness inference techniques (the simples one would be to use a similar technique as google.com uses, or used to use, for ranking websites).

\subsection{Notes on the privacy/anonymity challenge}
\label{sec:cha_privacy}

Depending on the subject, reporters may want to stay anonymous, e.g., in case of wikileaks.org.
Various methods have been established for that, with the existence of a middleman that keeps the identity of his/her source private being one of the simplest but most powerful.
More sophisticated methods are based on privacy-preserving approaches to machine learning or on cryptography.

Another important problem is that of protecting the privacy interests of people and (commercial or non-commerical) organizations mentioned in the entered reports, to the extent such privacy is reasonable.

\section{Conclusion}
\label{sec:conclusion}

We discussed, on an abstract level, various challenges and existing approaches for socio-economic transparency mechanisms.
From this rather superficial view, it seems that, so far, only individual points of a potential spectrum of socio-economical transparency mechanisms have been employed in practice.

Promising next steps are:
\begin{itemize}
\item Connecting more sensory information to transparency mechanisms: e.g. sensory information from cars, washing machines, etc. that are already in use to inform potential buyers about durability, energy consumption, emissions, etc. of these products.
\item Using a broader information basis, e.g., ``trustworthiness graphs'' as described above, to assess credibility of information sources.
\item Using machine learning techniques more extensively, in particular privacy-preserving methods, e.g.\ to learn how to filter out the relevant information based on a reward signal from users.
\end{itemize}

However, many questions remain open, e.g.\ in terms financing transparency mechanisms.

\bibliographystyle{plain}
\bibliography{include/transparency_mechanisms}

\end{document}